\documentclass{article}

\usepackage[preprint]{neurips_2024}

\usepackage[utf8]{inputenc} 
\usepackage[T1]{fontenc}    
\usepackage{hyperref}       
\usepackage{url}            
\usepackage{booktabs}       
\usepackage{amsfonts}       
\usepackage{nicefrac}       
\usepackage{microtype}      
\usepackage{xcolor}         

\usepackage{graphicx}
\usepackage{float} 
\usepackage{subfigure}
\usepackage{amsmath}
\usepackage{bbm}  
\usepackage{amssymb}
\usepackage{bm}
\usepackage{algorithm}
\usepackage{algpseudocode}
\usepackage{multirow}

\usepackage{appendix}
\usepackage{colortbl}

\usepackage{natbib}
\setcitestyle{numbers,square}

\usepackage{mathtools}

\usepackage{amsthm}  
\newtheorem{theorem}{Theorem}

\newtheorem{lemma}[theorem]{Lemma}

\newtheorem{definition}{Definition}[section]

\theoremstyle{remark}

\newcommand{\mysplit}[1]{%
  \begin{tabular}{@{}c@{}}   
    #1
  \end{tabular}
  }

\title{LFFR: Logistic Function For (multi-output) Regression}

\author{%
\href{https://orcid.org/0000-0003-0378-0607}{\includegraphics[scale=0.06]{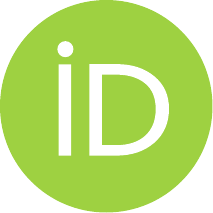}\hspace{1mm}John Chiang }
  \\
  \texttt{john.chiang.smith@gmail.com} \\
}

\begin{document}
\maketitle
\begin{abstract}
In this manuscript, we extend our previous work on privacy-preserving regression to address multi-output regression problems using data encrypted under a fully homomorphic encryption scheme. We build upon the simplified fixed Hessian approach for linear and ridge regression and adapt our novel LFFR algorithm, initially designed for single-output logistic regression, to handle multiple outputs. We further refine the constant simplified Hessian method for the multi-output context, ensuring computational efficiency and robustness. Evaluations on multiple real-world datasets demonstrate the effectiveness of our multi-output LFFR algorithm, highlighting its capability to maintain privacy while achieving high predictive accuracy. Normalizing both data and target predictions remains essential for optimizing homomorphic encryption parameters, confirming the practicality of our approach for secure and efficient multi-output regression tasks.
\end{abstract}

\section{Introduction}

\subsection{Background}
Multi-output regression~\cite{tsoumakas2014multi,borchani2015survey,spyromitros2016multi}, also referred to as multi-target, multi-variate, or multi-response regression, seeks to predict several real-valued output variables at the same time. When these output variables are binary, the problem is termed multi-label classification. Conversely, when the output variables are discrete (but not necessarily binary), it is known as multi-dimensional classification. Multi-target regression focuses on the simultaneous prediction of multiple continuous target variables using a single set of input variables. This approach is relevant in numerous industrial and environmental applications, including ecological modeling and energy forecasting.


Regression algorithms typically necessitate access to users' private data to construct accurate predictive models. This requirement raises substantial concerns about data privacy breaches, which can significantly deter users from sharing their sensitive information. To mitigate these privacy risks, various solutions have been proposed. Among these, homomorphic encryption (HE) is recognized as one of the most secure and promising approaches. HE enables computations to be performed directly on encrypted data without the need to decrypt it first, thereby preserving the privacy and confidentiality of the users' data throughout the computation process. This ensures that sensitive information remains protected, addressing the privacy concerns that often accompany the use of regression algorithms and encouraging users to share their data for more accurate model training.

In this paper, we extend a novel and efficient algorithm termed LFFR, initially designed for single-output regression using logistic regression-like neural network models,  to address multi-output regression problems. We aim to develop innovative algorithms that utilize the (Simplified) Fixed Hessian~\cite{IDASH2018bonte} method, enabling their effective application in encrypted states. This approach not only preserves data privacy but also enhances regression performance in non-linear scenarios.

\subsection{Related work}
Unlike privacy-preserving neural network inference~\cite{gilad2016cryptonets, li2018privacy, bourse2018fast, chiang2022volleyrevolver, chillotti2021programmable,kim2018matrix,chabanne2017privacy} and homomorphic logistic regression training~\cite{chiang2022privacy, kim2018logistic,kim2018secure,IDASH2018bonte,aono2016scalable,IDASH2018gentry,IDASH2019kim},, which have been extensively researched and developed with significant contributions from the iDASH competitions in 2018 and 2019, the area of regression learning without revealing private information has not received as much attention. Homomorphic encryption (HE)-based approaches and multi-party computation (MPC)-based approaches are two primary methods considered in this context. Each of these methods offers unique advantages in addressing the privacy concerns inherent in traditional regression learning algorithms. 

To the best of our knowledge, this work represents the first study on training multi-output regression models under homomorphic encryption.

\paragraph{HE-Based Approaches.}
Graepel et al.\cite{graepel2012ml} utilized Somewhat Homomorphic Encryption, which supports a limited number of multiplications, to solve a linear system in a privacy-preserving manner using gradient descent. Ogilvie et al.\cite{ogilvie2020improved} accomplished privacy-preserving ridge regression training through homomorphic encryption via the Simplified Fixed Hessian method.

We have developed a novel algorithm for non-linear regression tasks, which can be computed using a similar process to linear regression over transformed data.

\paragraph{MPC-Based Approaches.} Nikolaenko et al.\cite{nikolaenko2013privacy} introduced an MPC-based protocol for privacy-preserving linear regression training, combining linear homomorphic encryption with garbled circuit construction. Hall et al.\cite{hall2011secure} proposed another MPC-based protocol for training linear regression models, based on secret sharing under the semi-honest model.

However, MPC-based approaches often involve significant communication overhead and require that at most one of the two parties is malicious.

\subsection{Contributions}
Our primary contribution is the extension of the LFFR algorithm from single-output regression to a multi-output version, enabling privacy-preserving regression training using simplified fixed Hessian minimization. This approach is based on a simple 2-layer neural network architecture similar to logistic regression. We derived a simplified fixed Hessian for our multi-output LFFR algorithm and further enhanced it to eliminate the need for computing the sigmoid function, utilizing the same calculation circuit as linear regression. This improved version effectively models non-linear relationships between variables.

Additionally, we emphasize the importance of normalizing not only the input data but also the predictions when implementing regression algorithms in the encrypted domain. This practice offers several advantages and is highly recommended for achieving optimal performance.

\section{Preliminaries}


\subsection{Fully Homomorphic Encryption}
Homomorphic encryption is a class of encryption schemes that allows the evaluation of arithmetic circuits directly on ciphertexts without needing to decrypt them first. A particularly valuable variant within this class is Fully Homomorphic Encryption (FHE), which supports functions with unlimited multiplicative depth, thus enabling the computation of any function on encrypted data. Historically, FHE schemes faced significant challenges in managing the magnitude of plaintexts when attempting to apply FHE to real-world machine learning tasks. However, Cheon et al.~\cite{cheon2017homomorphic} introduced an efficient approach called \(\texttt{CKKS}\), incorporating a \(\texttt{rescaling}\) procedure that is suitable for handling practical data types such as real numbers.

The \(\texttt{CKKS}\) scheme supports approximate arithmetic on floating-point numbers, interpreting noise as part of the usual errors observed in cleartext when performing arithmetic operations in floating-point (fractional) representation. It features a native encoding function based on the canonical embedding and, like several other homomorphic encryption schemes, allows for the encoding of multiple messages into a single plaintext (known as \(\texttt{SIMD}\)), facilitating "slotwise" operations on the messages using the automorphism graph. The open-source library \(\texttt{HEAAN}\), which implements the \(\texttt{CKKS}\) scheme as presented by Cheon et al.~\cite{cheon2017homomorphic, cheon2018bootstrapping}, includes all the algorithms necessary to perform modern homomorphic operations.

We denote by \(\texttt{ct.m}\) a ciphertext that encrypts the message \(m\). For further details and additional information, readers are encouraged to refer to the works of Cheon et al.~\cite{cheon2017homomorphic, cheon2018bootstrapping}.

The application of the \(\texttt{SIMD}\) parallel technique has led to the development of numerous refined algorithms that are useful for calculating complex tasks, such as computing gradients.


\subsection{Simplified Fixed Hessian}
The Newton-Raphson method, commonly referred to as Newton's Method, is primarily used to find the roots of real-valued functions. However, it is computationally expensive due to the need to calculate and invert the Hessian matrix. To mitigate this, Böhning and Lindsay~\cite{bohning1988monotonicity} developed the Fixed Hessian Method, a variant of Newton's Method that reduces computational costs by utilizing a fixed (constant) approximation of the Hessian matrix rather than recalculating it at every iteration. In 2018, Bonte and Vercauteren~\citep{IDASH2018bonte} proposed the Simplified Fixed Hessian (SFH) method for logistic regression training over the BFV homomorphic encryption scheme. This method further simplifies the Fixed Hessian approach by constructing a diagonal Hessian matrix using the Gershgorin circle theorem.

Both the Fixed Hessian method and the Simplified Fixed Hessian method encounter issues when applied to high-dimensional sparse matrices, such as the MNIST datasets, due to the potential for a singular Hessian substitute. Additionally, the construction of the simplified fixed Hessian requires that all dataset elements be non-negative. If datasets are normalized to the range $[-1, +1]$, the SFH method may fail to converge, as it does not meet the convergence conditions of the Fixed Hessian method. These limitations were addressed by Chiang~\cite{chiang2022privacy}, who introduced a faster gradient variant called the $\texttt{quadratic gradient}$. This variant generalizes the SFH to be invertible under any circumstances.

Furthermore, the works of Böhning and Lindsay~\cite{bohning1988monotonicity} and Bonte and Vercauteren~\cite{IDASH2018bonte} do not provide a systematic approach for finding a (Simplified) Fixed Hessian substitute. Most objective functions may not possess a suitable constant bound for the Hessian matrix. Chiang presents a method to probe the existence of an SFH for the objective function, which is a necessary condition. This means that while a function might still have an SFH, Chiang's method may fail to identify it.

For any twice-differentiable function $F(\mathbf{x})$, if the Hessian matrix $H$ with elements $\bar{h}{ki}$ is positive definite for minimization problems (or negative definite for maximization problems), the SFH method first requires finding a constant matrix $\bar{H}$ that satisfies $\bar{H} \le H$ in the Loewner ordering for a maximization problem, or $H \le \bar{H}$ otherwise. The SFH can then be obtained in the following form:
\begin{equation*}
\begin{aligned}
B =
\left[ \begin{array}{cccc}
\sum_{i=0}^{d} \bar{h}_{0i} & 0 & \ldots & 0 \\
0 & \sum_{i=0}^{d} \bar{h}_{1i} & \ldots & 0 \\
\vdots & \vdots & \ddots & \vdots \\
0 & 0 & \ldots & \sum_{i=0}^{d} \bar{h}_{di} \\
\end{array}
\right].
\end{aligned}
\end{equation*}
Recall that the SFH method requires the dataset to have positive entries and that it may still be non-invertible.

Chiang's quadratic gradient~\cite{chiang2022privacy} also necessitates constructing a diagonal matrix $\tilde{B}$ in advance. However, unlike the SFH, this matrix does not need to be constant. It is derived from the Hessian matrix $H$ with entries $H_{0i}$:
\begin{equation*}
\begin{aligned}
\tilde{B} =
\left[ \begin{array}{ccc}
-\epsilon - \sum_{i=0}^{d} |H_{0i}| & 0 & \ldots \\
0 & -\epsilon - \sum_{i=0}^{d} |H_{1i}| & \ldots \\
\vdots & \vdots & \ddots \\
0 & 0 & \ldots \\
\end{array}
\right],
\end{aligned}
\end{equation*}
where $\epsilon$ is a small constant (typically $1e-8$) added to ensure numerical stability and to prevent division by zero. Notably, $\tilde{B}$ satisfies the convergence condition of the Fixed Hessian method, regardless of whether the dataset contains negative entries.




\subsection{Linear Regression Model}
We briefly review linear regression for multi-output regression.

Given a dataset matrix \( X \in \mathbb{R}^{n \times (1 + d)} \), where each row represents a record with \( d \) features plus an additional constant term (the first element being 1), and an outcome vector \( Y \in \mathbb{N}^{n \times 1} \) consisting of \( n \) corresponding predictions, let each record \( x_{[i]} \) have \( c \) observed predictions labeled as \( y_{i} \in \{ 1, 2, \ldots, c \} \). The multi-output linear regression model includes \( c \) parameter vectors \( w_{i} \), each of size \( (1 + d) \), which together form the parameter matrix \( W \). Each vector \( \mathbf{w}_{i} \) is used to model the \( i \)-th prediction.

\begin{align*}
  X &= 
 \begin{bmatrix}
 \textsl x_{1}      \\
 \textsl x_{2}      \\
 \vdots          \\
 \textsl x_{n}      \\
 \end{bmatrix}
 = 
 \begin{bmatrix}
 x_{[1][0]}    &   x_{[1][1]}   &  \cdots  & x_{[1][d]}   \\
 x_{[2][0]}    &   x_{[2][1]}   &  \cdots  & x_{[2][d]}   \\
 \vdots    &   \vdots   &  \ddots  & \vdots   \\
 x_{[n][0]}    &   x_{[n][1]}   &  \cdots  & x_{[n][d]}   \\
 \end{bmatrix},    \\
 Y &=  
 \begin{bmatrix}
 \textsl y_{1}     \\
 \textsl y_{2}     \\
 \vdots         \\
 \textsl y_{n}     \\
 \end{bmatrix}
  =
 \begin{bmatrix}
 y_{[1][1]}    &   y_{[1][2]}   &  \cdots  & y_{[1][c]}   \\
 y_{[2][1]}    &   y_{[2][2]}   &  \cdots  & y_{[2][c]}   \\
 \vdots    &   \vdots   &  \ddots  & \vdots   \\
 y_{[n][1]}    &   y_{[n][2]}   &  \cdots  & y_{[n][c]}   \\
 \end{bmatrix},  \\
 W &=  
 \begin{bmatrix}
 \textsl w_{1}     \\
 \textsl w_{2}     \\
 \vdots         \\
 \textsl w_{c}     \\
 \end{bmatrix}  =
 \begin{bmatrix}
 w_{[1][0]}    &   w_{[1][1]}   &  \cdots  & w_{[1][d]}   \\
 w_{[2][0]}    &   w_{[2][1]}   &  \cdots  & w_{[2][d]}   \\
 \vdots    &   \vdots   &  \ddots  & \vdots   \\
 w_{[c][0]}    &   w_{[c][1]}   &  \cdots  & w_{[c][d]}   \\
 \end{bmatrix}. 
\end{align*} 

Multi-output linear regression aims to find the best parameter matrix $W$ such that for each $i$ $$ y_{[i][1]}  \approx \textsl x_{i}^{\top}   \textsl w_{1},  y_{[i][2]}  \approx \textsl x_{i}^{\top}   \textsl w_{2},  \cdots , y_{[i][c]}  \approx \textsl x_{i}^{\top}   \textsl w_{c}, $$
which can be transformed into an optimization problem to minimize~\footnote{when applying first-order gradient descent methods, the common practise is to averge the squared residuals: $L_0 = \frac{1}{n}\sum_{i = 1}^{n} \frac{1}{d} \sum_{j = 1}^{d} (\textsl x_{i}^{\top}\textsl w_{j} - y_{[i][j]})^2$}:
$$L_0 = \sum_{i = 1}^{n} \sum_{j = 1}^{d} (\textsl x_{i}^{\top}\textsl w_{j} - y_{[i][j]})^2,$$ 
where $\mathbf{x}_i = [1 \ x_i]$ is the feature vector $x_i$ with a $1$ inserted at front by appending a 1 to each of the inputs. 

Multi-output linear regression solutions can be expressed in closed form as:
\[
\boldsymbol{\beta} = X^+ Y,
\]
where \( X^+ \) represents the Moore-Penrose pseudo-inverse of the matrix \( X \). The Moore-Penrose pseudo-inverse generalizes the concept of matrix inversion and offers a means to find the least-squares solution, which minimizes the Euclidean norm of the residuals \( \|X^\top W - Y\| \). 

To compute \( X^+ \), singular value decomposition (SVD) is commonly employed. However, this approach can be challenging to implement within the framework of fully homomorphic encryption (FHE). When obtaining the closed-form solution proves to be computationally prohibitive or infeasible, particularly with large datasets, alternative methods such as iterative optimization techniques—like gradient descent or Newton's method—can be utilized to directly minimize the cost function \( L_0 \).
The gradient and Hessian of the cost function $L_0(\boldsymbol{\beta})$ are given by, respectively:

\begin{align*}
 \nabla & = \frac{ \partial L }{  \partial W  } \\
        &= \Big [ \frac{ \partial  L }{  \partial \textsl{w}_{1} }, \frac{ \partial  L }{ \partial \textsl{w}_{2} } , \hdots, \frac{ \partial  L }{  \partial \textsl{w}_{c} }  \Big ]^{\top}  \\  
&= 
\Big [ \frac{ \partial  L }{  \partial {w}_{[1][0]} },\hdots,\frac{ \partial  L }{  \partial {w}_{[1][d]} },  \frac{ \partial  L }{  \partial {w}_{[2][0]} },\hdots,\frac{ \partial  L }{  \partial {w}_{[2][d]} }, \hdots, \frac{ \partial  L }{  \partial {w}_{[c][0]} },\hdots,\frac{ \partial  L }{  \partial {w}_{[c][d]} } \Big ]^{\top}  \\
&= \Big [ 2 \sum_i (\textsl x_i^{\top} \textsl w_{1} - y_{[i][1]} ) \textsl x_i,  2 \sum_i (\textsl x_i^{\top} \textsl w_{2} - y_{[i][2]} ) \textsl x_i, \cdots,2 \sum_i (\textsl x_i^{\top} \textsl w_{c} - y_{[i][c]} ) \textsl x_i \Big ]^{\top} 
\\
\\
 \nabla^2 & = \frac{ \partial^2 L }{  \partial W^2  } \\
 &=  
 \begin{bmatrix}
 \frac{ \partial^2  L }{  \partial \textsl{w}_{1} \partial \textsl{w}_{1} }    &   \frac{ \partial^2  L }{  \partial \textsl{w}_{1} \partial \textsl{w}_{2} }   &  \cdots  & \frac{ \partial^2  L }{  \partial \textsl{w}_{1} \partial \textsl{w}_{c} }   \\
 \frac{ \partial^2  L }{  \partial \textsl{w}_{2} \partial \textsl{w}_{1} }    &   \frac{ \partial^2  L }{  \partial \textsl{w}_{2} \partial \textsl{w}_{2} }   &  \cdots  & \frac{ \partial^2  L }{  \partial \textsl{w}_{2} \partial \textsl{w}_{c} }   \\
 \vdots    &   \vdots   &  \ddots  & \vdots   \\
 \frac{ \partial^2  L }{  \partial \textsl{w}_{c} \partial \textsl{w}_{1} }    &   \frac{ \partial^2  L }{  \partial \textsl{w}_{c} \partial \textsl{w}_{1} }   &  \cdots  & \frac{ \partial^2  L }{  \partial \textsl{w}_{c} \partial \textsl{w}_{c} }   \\
 \end{bmatrix}  \\
  &=  
 \begin{bmatrix}
2 \cdot \sum_i {\textsl{x}_i}^{\top}\textsl{x}_i   &   0   &  \cdots  &  0   \\
 0    &   2 \cdot \sum_i {\textsl{x}_i}^{\top}\textsl{x}_i    &  \cdots  &  0   \\
 \vdots    &   \vdots   &  \ddots  & \vdots   \\
 0    &    0   &  \cdots  & 2 \cdot \sum_i {\textsl{x}_i}^{\top}\textsl{x}_i    \\
 \end{bmatrix}  \\
 &=
 2I 
 \otimes
   [ X^{\top}X ],
\end{align*} 
where $\frac{ \partial^2 L }{  \partial \textsl{w}_{i} \partial \textsl{w}_{j} }  = \frac{ \partial^2 L }{  \partial \textsl{w}_{j} \partial \textsl{w}_{i} }  =   2 \cdot \sum_i {\textsl{x}_i}^{\top}\textsl{x}_i$ for $i = j$ and $\frac{ \partial^2 L }{  \partial \textsl{w}_{i} \partial \textsl{w}_{j} }  = \frac{ \partial^2 L }{  \partial \textsl{w}_{j} \partial \textsl{w}_{i} }  = 0 $ for $i \ne j$ and ``$\otimes$'' is the kronecker product.

The Kronecker product, unlike standard matrix multiplication, is an operation that combines two matrices of arbitrary sizes to produce a block matrix. For an $m \times n$ matrix $A$ and a $p \times q$ matrix $B$, the Kronecker product of $A$ and $B$ results in a $pm \times qn$ block matrix, denoted by $A \otimes B$. 

The inverse of the Kronecker product $A \otimes B$ exists if and only if both $A$ and $B$ are invertible. When both matrices are invertible, the inverse of the Kronecker product can be expressed as:

\begin{align}
   (A \otimes B)^{-1} = A^{-1} \otimes B^{-1}. \label{KP invertible product property} 
\end{align}

This property of the Kronecker product can be utilized to simplify the computation of the inverse of the Hessian matrix ${\nabla^2}$.

We strongly recommend normalizing the regressand to the range $[0, 1]$ or $[-1, +1]$. This normalization is beneficial in mitigating overfitting by constraining the range of the coefficients, which also simplifies the refresh of ciphertext due to the uniform coefficient range. Additionally, this approach helps avoid the need for parameter tuning, such as adjusting $\lambda$. Since ridge regression is conceptually similar to linear regression, with ridge regression being a regularized version of linear regression, we focus exclusively on linear regression in this paper and do not delve into ridge regression.

\section{Technical Details}
In this section, we first give our SFH for multi-output linear regression, then present our new algorithm for  multi-output regression, and finally descript its improved version.

\subsection{Our Linear Regression Algorithm}
\label{ alg:our LR algorithm }

To apply the Simplified Fixed Hessian method to the multi-output linear regression, we need to build the Simplified Fixed Hessian matrix $\bar B$ first, trying to find a fixed Hessian matrix. We can try to build it directly from the Hessian matrix $\nabla^2$ based on the construction method of quadratic gradient:
\begin{align*}
 \bar B &=  
  \begin{bmatrix}
 \bar B_{[0]} & \mathbf{0} & \cdots & \mathbf 0   \\
 \mathbf 0        & \bar B_{[1]} & \cdots & \mathbf 0   \\
 \vdots  & \vdots & \ddots & \vdots       \\
 \mathbf 0 & \mathbf 0 & \cdots & \bar B_{[c-1]}   \\
 \end{bmatrix} 
 \Big (  \bar B_{[i]} =  
  \begin{bmatrix}
 \bar B_{[i][0]} & 0 & \cdots &  0   \\
 0        & \bar B_{[i][1]} & \cdots & 0   \\
 \vdots  & \vdots & \ddots & \vdots       \\
  0 & 0 & \cdots & \bar B_{[i][d]}   \\
 \end{bmatrix}  \Big )  
  \\
 &=
diag\{\bar B_{[0][0]},\bar B_{[0][1]},\cdots,\bar B_{[0][d]},\bar B_{[1][0]},\bar B_{[1][1]},\cdots,\bar B_{[1][d]},\cdots,\bar B_{[c-1][0]},\bar B_{[c-1][1]},\cdots,\bar B_{[c-1][d]} \} ,
\end{align*} 
where
$ \bar B_{[i][j]} =
 2 \times (|x_{[1][j]}x_{[1][0]}|+\hdots+|x_{[1][j]}x_{[1][d]}|)   + \epsilon .$

Böohning~\citep{bohning1992multinomial} presented another lemma about the Kronecker product: If \( A \le B \) in the Loewner order, then
\begin{align}
A \otimes C \le B \otimes C \label{Bohning inequality property}
\end{align}
for any symmetric, nonnegative definite \( C \).

The Hessian for the cost function \( L_0 \) is already "fixed" but can still be singular, which can be corrected by applying a quadratic gradient. Following the construction method, from the Hessian \( H \) we can obtain a diagonal matrix \( \bar{B} \) with diagonal elements:
\[ \bar{B}_{kk} = \epsilon + 2 \sum_{j=0}^d \sum_{i=1}^n |x_{ij} x_{ik}|, \]
which has its inverse \( \bar{B}^{-1} \) in any case. The inverse \( \bar{B}^{-1} \) is also a diagonal matrix with diagonal elements \( \frac{1}{ \bar{B}_{kk} } \).

Replacing the full Hessian \( H \) for \( L_0 \) with our diagonal matrix \( \bar{B} \), updates now become:
\[ \boldsymbol{\beta} \leftarrow \boldsymbol{\beta} - \bar{B}^{-1} \nabla_{\boldsymbol{\beta}} L_0(\boldsymbol{\beta}). \]

\subsection{Logistic Function for Regression}
Like the LFFR algorithm for single-output regression, our multi-output version for LFFR also normalizes each predictions into the range $[0, 1]$ and records each minimum and maximum for each prediction.
 
Seen as an extension of multi-output linear regression, our algorithm aims to minimize the following cost function $L2$, similar to $L_0$: 
$$L_2 = \sum_{i = 1}^{n} \sum_{j = 1}^{d} (\sigma( \textsl x_{i}^{\top}\textsl w_{j}) - y_{[i][j]})^2.$$

The function $L_2$ has its gradient $g_2$ and Hessian $H_2$, given by:

\begin{align*}
 \nabla & = \frac{ \partial L_2 }{  \partial W  } \\
        &= \Big [ \frac{ \partial  L_2 }{  \partial \textsl{w}_{1} }, \frac{ \partial  L_2 }{ \partial \textsl{w}_{2} } , \hdots, \frac{ \partial  L_2 }{  \partial \textsl{w}_{c} }  \Big ]^{\top}  \\  
&= 
\Big [ \frac{ \partial  L_2 }{  \partial {w}_{[1][0]} },\hdots,\frac{ \partial  L_2 }{  \partial {w}_{[1][d]} },  \frac{ \partial  L_2 }{  \partial {w}_{[2][0]} },\hdots,\frac{ \partial  L_2 }{  \partial {w}_{[2][d]} }, \hdots, \frac{ \partial  L_2 }{  \partial {w}_{[c][0]} },\hdots,\frac{ \partial  L_2 }{  \partial {w}_{[c][d]} } \Big ]^{\top}  \\
&= \Big [ 2 \sum_i (\sigma(i1) - y_{[i][1]} )\sigma(i1)( 1 - \sigma(i1))\textsl x_i,  2 \sum_i (\sigma(i2) - y_{[i][2]} )\sigma(i2)( 1 - \sigma(i2))\textsl x_i, \cdots,  2 \sum_i (\sigma(ic) - y_{[i][c]} )\sigma(ic)( 1 - \sigma(ic))\textsl x_i \Big ]^{\top} 
\\
\\
 \nabla^2 & = \frac{ \partial^2 L_2 }{  \partial W^2  } \\
 &=  
 \begin{bmatrix}
 \frac{ \partial^2  L_2 }{  \partial \textsl{w}_{1} \partial \textsl{w}_{1} }    &   \frac{ \partial^2  L_2 }{  \partial \textsl{w}_{1} \partial \textsl{w}_{2} }   &  \cdots  & \frac{ \partial^2  L_2 }{  \partial \textsl{w}_{1} \partial \textsl{w}_{c} }   \\
 \frac{ \partial^2  L_2 }{  \partial \textsl{w}_{2} \partial \textsl{w}_{1} }    &   \frac{ \partial^2  L_2 }{  \partial \textsl{w}_{2} \partial \textsl{w}_{2} }   &  \cdots  & \frac{ \partial^2  L_2 }{  \partial \textsl{w}_{2} \partial \textsl{w}_{c} }   \\
 \vdots    &   \vdots   &  \ddots  & \vdots   \\
 \frac{ \partial^2  L_2 }{  \partial \textsl{w}_{c} \partial \textsl{w}_{1} }    &   \frac{ \partial^2  L_2 }{  \partial \textsl{w}_{c} \partial \textsl{w}_{1} }   &  \cdots  & \frac{ \partial^2  L_2 }{  \partial \textsl{w}_{c} \partial \textsl{w}_{c} }   \\
 \end{bmatrix}  \\
  &=  
 \begin{bmatrix}
\Delta_i \cdot \sum_i {\textsl{x}_i}^{\top}\textsl{x}_i   &   0   &  \cdots  &  0   \\
 0    &   \Delta_i \cdot \sum_i {\textsl{x}_i}^{\top}\textsl{x}_i    &  \cdots  &  0   \\
 \vdots    &   \vdots   &  \ddots  & \vdots   \\
 0    &    0   &  \cdots  & \Delta_i \cdot \sum_i {\textsl{x}_i}^{\top}\textsl{x}_i    \\
 \end{bmatrix}  \\
 &=
 \Delta_i 
 \otimes
   [ X^{\top}X ],
\end{align*} 
where  $\sigma(ij) = \sigma(\textsl x_{i}^{\top}   \textsl w_{j}),$ $\Delta_i = (4\sigma(i) - 6\sigma(i)^2 - 2y_i + 4y_i\sigma(i))\sigma(i)(1 - \sigma(i)),$ and $S$ is a diagonal matrix with diagonal entries $\Delta_i$.

According to LFFR for single-output regression, we obtain the final maximum value $\Delta_i \le 0.155$.

So far, we compute our simplified fixed Hessian $\bar B$ for $L_2.$ That is, a diagonal matrix with diagonal elements $\bar B_{kk}$ shown as:
$$\bar B_{kk} = \epsilon + 0.155\sum_{j = 0}^d \sum_{i = 1}^n |x_{ij}x_{ik}|.$$ Our updates for LFFR are now given component wise by:
\begin{align}
\beta_j \leftarrow \beta_j + \frac{1}{\bar B_{jj}} \nabla_{\boldsymbol{\beta}} L_2(\boldsymbol{\beta}).
\label{ eq:LFFR }
\end{align}

Given the input features \( X \) and the target values \( \mathbf{y} \), our revised LFFR algorithm incorporating the normalization for prediction values consists of the following three steps:

\indent \texttt{Step 1:}
We preprocess the dataset to ensure \( X \) and \( \mathbf{y} \) are both in the required format. If the regression task is to predict continuous values other than likelihood, we normalize the target values to fall within the \([0, 1]\) range. For example, we can iterate through the predicted values \( \mathbf{y} \) to find the minimum \((y_{\min})\) and maximum \((y_{\max})\) values and scale each prediction \( y_i \) to a common range \([0, 1]\) by the following formula:
\[ \bar{y}_i = \frac{y_i - y_{\min}}{y_{\max} - y_{\min} + \epsilon}, \]
where \( \bar{y}_i \) is the normalized value of \( y_i \).
The training data \( X \) should also be scaled to a fixed range, usually \([0, 1]\) or \([-1, 1]\).

\indent \texttt{Step 2:}
Using the data \( X \) and the new predictions \( \bar{\mathbf{y}} \) consisting of normalized values \( \bar{y}_i \), we train the LFFR model to fit the transformed dataset, updating the model parameter vector \( \boldsymbol{\beta} \) by the formula~(\ref{ eq:LFFR }). After the completion of the training process, the trained model weights \( \boldsymbol{\beta}_{\text{trained}} \) are obtained.

\indent \texttt{Step 3:}
When a new sample \( \mathbf{x} \) arrives, we first apply the same normalization technique used on the training data to \( \mathbf{x} \), resulting in a normalized sample \( \bar{\mathbf{x}} \). Then, \( \boldsymbol{\beta}_{\text{trained}} \) is employed to compute the new sample's probability \( \text{prob} = \sigma(\boldsymbol{\beta}_{\text{trained}}^{\top} \bar{\mathbf{x}}) \). Finally, we utilize the inverse function of the normalization to map the likelihood \( \text{prob} \) back to continuous values, obtaining the prediction \( y_i \) in the original continuous space using \( y_{\min} \) and \( y_{\max} \):
\[ y_i = (y_{\max} - y_{\min} + \epsilon) \cdot \text{prob} + y_{\min}. \]

\subsection{The Improved LFFR Version}
After converting the contiunous predictions $\mathbf{y}$ into likilihoods $\bar{ \mathbf{y} }$, our LFFR algorihtm aims to solve a system of non-linear equaltions:
\begin{align}
\begin{cases}
   \sigma(\textsl x_1^{\top} \textsl w_{1}) \approx \bar{y}_{[1][1]} = \frac{y_{[1][1]} - y_{\min}}{y_{\max} - y_{\min} + \epsilon}, \cdots,   
      \sigma(\textsl x_1^{\top} \textsl w_{c}) \approx \bar{y}_{[1][c]} = \frac{y_{[1][c]} - y_{\min}}{y_{\max} - y_{\min} + \epsilon},   \\
   \sigma(\textsl x_2^{\top} \textsl w_{1}) \approx \bar{y}_{[2][1]} = \frac{y_{[2][1]} - y_{\min}}{y_{\max} - y_{\min} + \epsilon}, \cdots,   
      \sigma(\textsl x_2^{\top} \textsl w_{c}) \approx \bar{y}_{[2][c]} = \frac{y_{[2][c]} - y_{\min}}{y_{\max} - y_{\min} + \epsilon},   \\
   \hspace{3cm} \vdots \\
   \sigma(\textsl x_n^{\top} \textsl w_{1}) \approx \bar{y}_{[n][1]} = \frac{y_{[n][1]} - y_{\min}}{y_{\max} - y_{\min} + \epsilon}, \cdots,   
      \sigma(\textsl x_n^{\top} \textsl w_{c}) \approx \bar{y}_{[n][c]} = \frac{y_{[n][c]} - y_{\min}}{y_{\max} - y_{\min} + \epsilon}.
\end{cases}
\label{ eq:non-linear system }
\end{align}

It is important to point out that the converting function to normalize predictions could just be a linear bijective function rather than something else. For instance, if we use the sigmoid function to map $y_i$ into the range $[0, 1]$, our LFFR algorithm will reduce to the original linear regression since $\sigma(\boldsymbol{\beta}^{\top}\mathbf{x}_i) \approx \sigma(y_i) \Rightarrow \boldsymbol{\beta}^{\top}\mathbf{x}_i \approx y_i$ is the formula of linear regression.

The system of equations~$($\ref{ eq:non-linear system }$)$ can be transformed into:
\begin{align}
\begin{cases}
   \sigma^{-1}\sigma(\textsl x_1^{\top} \textsl w_{1}) = \textsl x_1^{\top} \textsl w_{1} \approx \sigma^{-1}(\bar{y}_{[1][1]}) , \cdots,   
      \sigma^{-1}\sigma(\textsl x_1^{\top} \textsl w_{c}) = \textsl x_1^{\top} \textsl w_{c} \approx \sigma^{-1}(\bar{y}_{[1][c]}),   \\
   \sigma^{-1}\sigma(\textsl x_2^{\top} \textsl w_{1}) = \textsl x_2^{\top} \textsl w_{1} \approx \sigma^{-1}(\bar{y}_{[2][1]}) , \cdots,   
      \sigma^{-1}\sigma(\textsl x_2^{\top} \textsl w_{c}) = \textsl x_2^{\top} \textsl w_{c} \approx \sigma^{-1}(\bar{y}_{[2][c]}),   \\
   \hspace{3cm} \vdots \\
   \sigma^{-1}\sigma(\textsl x_n^{\top} \textsl w_{1}) = \textsl x_n^{\top} \textsl w_{1} \approx \sigma^{-1}(\bar{y}_{[n][1]}) , \cdots,   
      \sigma^{-1}\sigma(\textsl x_n^{\top} \textsl w_{c}) = \textsl x_n^{\top} \textsl w_{c} \approx \sigma^{-1}(\bar{y}_{[n][c]}).
\end{cases}
\label{ eq:non-linear system transformed }
\end{align}
where $\sigma^{-1}(y) = \ln(\frac{y}{1-y}) $ for $0 \le y \le 1$ is the inverse of the sigmoid function, known as the logit function. 

The system of equations~(\ref{ eq:non-linear system transformed }) can be seen as linear regression over a new dataset consisting of the original samples and the new likelihood predictions converted from the original ones $y_i$ by $\sigma^{-1}\left(\frac{y_i - y_{\min}}{y_{\max} - y_{\min} + \epsilon}\right)$, which can be solved via our linear regression algorithm in Section~\ref{ alg:our LR algorithm }. Since $\sigma^{-1}(0)$ and $\sigma^{-1}(1)$ are negative infinity and positive infinity, respectively, for numerical stability we cannot normalize the original predicted values into the range containing $0$ and $1$. Therefore, we introduce a parameter $\gamma$ to control the range of the newly generated predictions within the interval $[0, 1]$, excluding $0$ and $1$, where $0 < \gamma < 1$. The new range for probability predictions will be $[0.5 - \gamma/2, 0.5 + \gamma/2]$.

Our improved LFFR algorithm, which removes the limitations of the original LFFR and is described in Algorithm~\ref{ alg:our improved LFFR algorithm }, consists of the following three steps:

\textbf{Step 1:}  
Whether or not the task is to predict likelihood, we normalize the prediction $y_i$ first and transform it to a new one $\bar{y}_i$, with the parameter $\gamma$ by the mapping: $\bar{y}_i = \sigma^{-1}\left(\frac{y_i - y_{\min}}{y_{\max} - y_{\min} + \epsilon} \cdot \gamma + 0.5 - \gamma/2\right)$.  
We assume that the dataset $X$ has already been normalized into a certain range like $[-1, +1]$. The normalized data $x_i$ with the new transformed predictions $\bar{y}_i$ is composed of a new regression task to predict some simulation likelihood, which can be solved by our linear regression algorithm in Section~\ref{ alg:our LR algorithm }.

\textbf{Step 2:}  
We apply our linear regression algorithm to address the new regression task of predicting simulation likelihoods $\bar{y}_i$, obtaining a well-trained weight vector $\boldsymbol{\beta}$. This step is the same as that of our LFFR algorithm.

\textbf{Step 3:}  
For some new, already normalized sample $\mathbf{x}$ to predict, we first use the well-trained weight vector $\boldsymbol{\beta}$ to compute its simulation likelihood $prob = \sigma(\boldsymbol{\beta}^{\top} \bar{\mathbf{x}})$. Note that $prob \in [0, 1]$ could exceed the range $[0.5 - \gamma/2, 0.5 + \gamma/2]$ and therefore let our improved LFFR algorithm predict continuous values outside the original prediction range $[y_{\min}, y_{\max}]$. 

Then, we utilize the inverse function of the new normalization mapping to convert the likelihood $prob$ back to the true prediction value $\bar{y}$ in the original continuous space using $y_{\min}$ and $y_{\max}$:
$$\bar{y} = (y_{\max} - y_{\min} + \epsilon) \cdot \frac{prob - 0.5 + \gamma/2}{\gamma} + y_{\min}.$$

\begin{algorithm}[htbp]
    \caption{Our Improved LFFR algorithm }
    \label{ alg:our improved LFFR algorithm  }
     \begin{algorithmic}[1]
        \Require Normalized dataset $ X \in \mathbb{R} ^{n \times (1+d)} $, target values $ Y \in \mathbb{R} ^{n \times 1} $, float number parameter $ \gamma \in \mathbb{R} (\texttt{set to 0.5 in this work}) $, and the number  $\kappa$ of iterations
        \Ensure the weight vector $ \boldsymbol{\beta} \in \mathbb{R} ^{(1+d)} $ 

        \State Set $ \boldsymbol{\beta} \gets \boldsymbol 0$
        \Comment{$\boldsymbol{\beta} \in \mathbb{R}^{(1+d)}$}        
        
        \State Set $\bar B \gets \boldsymbol 0$
        \State Set $\bar H \gets 2X^{\top}X$
        \Comment{$\bar H \in \mathbb{R}^{(1+d) \times (1+d)}$}
                
           \For{$i := 0$ to $d$}
              \State $\bar B[i][i] \gets \epsilon$
              \Comment{$\epsilon$ is a small positive constant such as $1e-8$}
              \For{$j := 0$ to $d$}
                 \State $ \bar B[i][i] \gets \bar B[i][i] + |\bar H[i][j]| $
              \EndFor
              \State $ \bar B[i][i] \gets 1 / \bar B[i][i] $
           \EndFor
       
        \State Set $ \bar Y \gets \boldsymbol 0$
        \Comment{$\boldsymbol{\beta} \in \mathbb{R}^{(1+d)}$} 
        \State Set $ Y_{\min} \gets \min(Y) $
        \State Set $ Y_{\max} \gets \max(Y) $
        \For{$i := 0$ to $d$}
           \State $ \bar Y[i] \gets  \sigma^{-1}( \frac{Y[i] - Y_{\min}}{Y_{\max} - Y_{\min} + \epsilon}*\gamma + 0.5 - \gamma/2) $
        \EndFor                 
                 
        \For{$k := 1$ to $\kappa$}
           \State Set $Z \gets \boldsymbol 0 $
           \Comment{$Z \in \mathbb{R}^{n}$  will store  the dot products     }
           \For{$i := 1$ to $n$}
              \For{$j := 0$ to $d$}
                 \State $ Z[i] \gets  Z[i] +  \boldsymbol{\beta}[j] \times  X[i][j] $ 
              \EndFor
           \EndFor

           \State Set $\boldsymbol g \gets \boldsymbol 0$
           \For{$j := 0$ to $d$}
              \For{$i := 1$ to $n$}
                 \State $\boldsymbol g[j] \gets \boldsymbol g[j] + 2 \times (Z[i] - \bar Y[i]) \times X[i][j] $
              \EndFor
           \EndFor
           
           \For{$j := 0$ to $d$}
              \State $ \boldsymbol{\beta}[j] \gets \boldsymbol{\beta}[j]  -  \bar B[j][j] \times \boldsymbol g[j] $
           \EndFor
           
        \EndFor
      
        \State \Return $ \boldsymbol{\beta} $
        \Comment{For a new normalized sample $\mathbf x$, The algorithm output the prediction: $(\sigma(\boldsymbol{\beta}^{\top}\mathbf{x}) - 0.5 + \gamma/2)/\gamma(Y_{\max} - Y_{\min} + \epsilon) + Y_{\min}$ }
        \end{algorithmic}
\end{algorithm}

\subsection{The Algorithm Adaptation Method}
Algorithm adaptation techniques, also known as global or big-bang approaches, involve modifying single-output methods (like decision trees and support vector machines) to handle multi-output datasets directly. These approaches are considered more complex because they not only strive to predict multiple targets simultaneously but also aim to capture and interpret the interdependencies among these targets.

\subsection{Comparison in The Clear}

We assess the performance of four algorithms: our LFFR algorithm (denoted as LFFR), our improved LFFR algorithm (ImprovedLFFR), our linear regression algorithm (LR), and its variant with normalized predictions (YnormdLR), in the clear setting using the Python programming language. To evaluate convergence speed, we choose the mean squared error (MSE) loss function during the training phase as the sole indicator.

Random datasets with different parameters are generated to evaluate these four algorithms. Let the data have $n$ samples $x_i$ normalized into the range $[-1, +1]$, with $d$ features. $d$ floating-point numbers $r_i$ from the range $[-1, +1]$ are randomly generated to represent the linear relationship between the sample and its prediction, with a noise $nois$ generated following a Gaussian distribution $\mathcal{N}(0, \sigma^2)$ with a mean of $0$ and a variance of $\sigma^2$. Therefore, the prediction $y_i$ for $x_i$ is given by $y_i = nois + \sum_{k = 1}^d r_k x_{ik}$. Our linear regression variant will normalize the predictions in advance, record the minimum and maximum values of the predictions, and use them in the inference phase to compute the real predictions in the original space.

\textbf{Analysis and Discussion} 



\section{Secure Regression }
When employed in the encrypted domain, our LFFR algorithm involving the calculation of the logistic function faces a significant challenge: no homomorphic encryption schemes are currently capable of directly calculating the sigmoid function. A common solution is to replace the sigmoid function with a polynomial approximation. Several works have been conducted to find such polynomials, and the current state-of-the-art technique~\cite{cheon2022homomorphicevaluation} can approximate the sigmoid function over fairly large intervals with reasonable precision, such as over the range $[-3000, +3000]$. In our implementation, we use a function named ``\texttt{polyfit($\cdot$)}'' in the Python package Numpy, utilizing the least squares approach to approximate the sigmoid function over the interval $[-5, 5]$ by a degree 3 polynomial:
\[ g(x) = 0.5 + 0.19824 \cdot x - 0.0044650 \cdot x^3 .\]

The challenge in applying the quadratic gradient is to invert the diagonal matrix $\tilde{B}$ in order to obtain $\bar{B}$. We delegate the computation of the matrix $\bar{B}$ to the data owner, who then uploads the ciphertext encrypting $\bar{B}$ to the cloud. Since the data owner is already preparing and normalizing the dataset, it is also feasible for the data owner to calculate $\bar{B}$ without leaking sensitive data information.

\subsection{Database Encoding Method} \label{basic he operations} 

Given the training dataset $\mathbf{X} \in \mathbb{R}^{n \times (1+d)}$ and training labels $\mathbf{Y} \in \mathbb{R}^{n \times 1}$, we adopt the same method that Kim et al.~\cite{kim2018logistic} used to encrypt the data matrix, which consists of the training data combined with training-label information into a single ciphertext $\text{ct}_Z$. The weight vector $\boldsymbol{\beta}^{(0)}$ consisting of zeros and the diagonal elements of $\bar{B}$ are copied $n$ times to form two matrices. The data owner then encrypts the two matrices into two ciphertexts $\text{ct}_{\boldsymbol{\beta}}^{(0)}$ and $\text{ct}_{\bar{B}}$, respectively. The ciphertexts $\text{ct}_Z$, $\text{ct}_{\boldsymbol{\beta}}^{(0)}$, and $\text{ct}_{\bar{B}}$ are as follows:

\begin{equation*}
 \begin{aligned}
\text{ct}_X &= Enc
\left[ \begin{array}{cccc}
 1  &   x_{11}  &  \ldots  &  x_{1d}  \\
 1  &   x_{21}  &  \ldots  &  x_{2d}  \\
 \vdots         &   \vdots         &  \ddots  &  \vdots         \\
 1  &  x_{n1}   &  \ldots  &  x_{nd}  \\
 \end{array}
 \right], 
&\text{ct}_Y = Enc
\left[ \begin{array}{cccc}
 y_{11}  &   y_{12}  &  \ldots  &  y_{1c}  \\
 y_{21}  &   y_{22}  &  \ldots  &  y_{2c}  \\
 \vdots         &   \vdots         &  \ddots  &  \vdots         \\
 y_{n1}  &   y_{n2}  &  \ldots  &  y_{nc}  \\
 \end{array}
 \right],   \\
 \text{ct}_{\boldsymbol{\beta}}^{(0)} &= Enc
\left[ \begin{array}{cccc}
 \beta_{0}^{(0)}  &   \beta_{1}^{(0)} &  \ldots  &  \beta_{d}^{(0)} \\
 \beta_{0}^{(0)}  &   \beta_{1}^{(0)} &  \ldots  &  \beta_{d}^{(0)}  \\
 \vdots         &   \vdots         &  \ddots  &  \vdots         \\
 \beta_{0}^{(0)}  &   \beta_{1}^{(0)} &  \ldots  &  \beta_{d}^{(0)}  \\
 \end{array}
 \right],  
& \text{ct}_{\bar B} = Enc
\left[ \begin{array}{cccc}
  \bar B_{[0][0]}    & \bar B_{[1][1]}  &  \ldots  & \bar B_{[d][d]}  \\
  \bar B_{[0][0]}    & \bar B_{[1][1]}  &  \ldots  & \bar B_{[d][d]}  \\
 \vdots  & \vdots                & \ddots  & \vdots     \\
  \bar B_{[0][0]}    & \bar B_{[1][1]}  &  \ldots  & \bar B_{[d][d]}  \\
 \end{array}
 \right], 
 \end{aligned}
\end{equation*}
where $\bar B_{[i][i]}$  is the diagonal element of  $\bar B$.

\subsection{The Usage Scenarios} 
Supposing two different roles including data oweners (e.g. hospital or individuals) and cloud computing service providers (e.g. Amazon, Google or Microsoft), our proposed algorithms can be employed to two main situations:

\subsection{The Whole Pipeline} 
The full pipeline of privacy-preserving logistic regression training consists of the following steps:

\indent \texttt{Step 1:} 
The client prepares two data matrices, $\bar{B}$ and $Z$, using the training dataset.

\indent \texttt{Step 2:} 
The client then encrypts three matrices, $\bar{B}$, $Z$, and the initial weight matrix $W^{(0)}$, into three ciphertexts $\text{ct}_Z$, $\text{ct}_{W}^{(0)}$, and $\text{ct}_{\bar{B}}$, using the public key provided by a third party under the assigned homomorphic encryption (HE) system. We recommend using a zero matrix as the initial weight matrix if an already-trained weight matrix is not available.

\indent \texttt{Step 3:} 
The client uploads the ciphertexts to the cloud server and completes its part of the process, awaiting the result.

\indent \texttt{Step 4:} 
The public cloud begins by evaluating the gradient ciphertext $\text{ct}_{g}^{(0)}$ using $\text{ct}_Z$ and $\text{ct}_{W}^{(0)}$ through various homomorphic operations. Kim et al.~\cite{kim2018logistic} provide a detailed description of the homomorphic evaluation of the gradient descent method.

\indent \texttt{Step 5:} 
The public cloud computes the quadratic gradient with a homomorphic multiplication of $\text{ct}_{\bar{B}}$ and $\text{ct}_{g}^{(0)}$, resulting in the ciphertext $\text{ct}_{G}^{(0)}$.

\indent \texttt{Step 6:} 
The weight ciphertext is updated with the quadratic gradient ciphertext using our enhanced mini-batch NAG method. This step consumes some modulus level of the weight ciphertext.

\indent \texttt{Step 7:} 
The public cloud checks if the remaining modulus level of the weight ciphertext allows for another round of updates. If not, the algorithm bootstraps the weight ciphertexts using some public keys, obtaining a new ciphertext with a larger modulus while encrypting the same weight.

\indent \texttt{Step 8:} 
The public cloud completes all iterations of homomorphic logistic regression training, obtains the resulting weight ciphertext, and returns a ciphertext encrypting the updated model vector to the client.

The client can then decrypt the received ciphertext using the secret key to obtain the logistic regression model for their exclusive use.

Han et al.~\cite{han2018efficient} provide a detailed description of their HE-friendly logistic regression algorithm with an HE-optimized iteration loop in the encrypted domain. Our enhanced mini-batch NAG method is similar to theirs, except it requires one more ciphertext multiplication between the gradient ciphertext and the uploaded ciphertext encrypting $\bar{B}_i$ for each mini-batch. For more information, please refer to~\cite{han2018efficient}.


The public cloud processes the three ciphertexts $\text{ct}_Z$, $\text{ct}_{\boldsymbol{\beta}}^{(0)}$, and $\text{ct}_{\bar B}$ to evaluate the enhanced Nesterov Accelerated Gradient (NAG) algorithm, which aims to find an optimal weight vector by iteratively updating $\text{ct}_{\boldsymbol{\beta}}^{(0)}$. For a detailed explanation on how to compute the gradient using homomorphic encryption programming, please refer to \cite{kim2018logistic}.

\section{Experiments}
\paragraph{Implementation}  We employ our linear regression algorithm with normalized predictions, along with our LFFR and Improved LFFR algorithms, to implement privacy-preserving regression training based on homomorphic encryption using the \texttt{HEAAN} library. The reason we do not implement linear regression without normalizing the predictions is due to the difficulty in setting varying HE parameters for different regression tasks. The C++ source code for our implementations is publicly available at \href{https://github.com/petitioner/ML.LFFR}{https://github.com/petitioner/ML.LFFR}. All the experiments on the ciphertexts were conducted on a public cloud with $1$ vCPU and $50$ GB RAM.


\paragraph{Datasets}
Our experiments are based on 12 datasets. Table 1 provides detailed information about these datasets, including the name (1st column), abbreviation (2nd column), and source (3rd column). It also reports the number of instances in the training and test sets, or the total number of instances if cross-validation was used (4th column), the number of input variables $p$ (5th column), and the number of output variables $q$ (6th column). These datasets have been chosen to comprehensively evaluate the performance of our proposed algorithm across various scenarios.

\paragraph{Parameters}
While the datasets involved in our experiments differ in size, we can still use the same \texttt{HEAAN} parameters for both: $\log N = 16$, $\log Q = 1200$, $\log p = 30$, and $slots = 32768$, which ensure a security level of $\lambda = 128$. For further details on these parameters, please refer to \cite{kim2018logistic}. Under this setting, the Boston Housing dataset is encrypted into a single ciphertext, whereas the California Housing dataset is encrypted into $33$ ciphertexts. Since our LFFR algorithm needs to compute the sigmoid function in the encrypted state, it consumes more modulus and requires refreshing its weight ciphertext every few iterations. In contrast, our improved LFFR algorithm requires this refresh less frequently. For our improved LFFR algorithm, selecting the optimum parameter $\gamma$ is beyond the scope of this work, and we set it to $0.5$ for simplicity. Our LFFR algorithm uses a degree $3$ polynomial $g(x)$ to approximate the sigmoid function.

\paragraph{Results}

\paragraph{Microbenchmarks}

\paragraph{Discussion}
Our algorithm is expected to exhibit improved performance in terms of convergence speed if we apply a quadratic gradient with an optimized learning rate configuration to the simplified fixed Hessian (SFH). By carefully tuning the learning rate and leveraging the quadratic gradient, we can enhance the efficiency of the optimization process, potentially leading to faster convergence and more accurate predictions. This approach can be particularly beneficial when dealing with large-scale datasets and complex regression tasks, where traditional methods may struggle with convergence and computational efficiency.

\section{Conclusion}

In this paper, we presented the multi-output version of an efficient algorithm named LFFR, which utilizes the logistic function. This algorithm can capture more complex relationships between input values and output predictions compared to traditional linear regression. Our extended LFFR algorithm is designed to handle multiple outputs simultaneously, making it highly suitable for practical applications where multiple dependent variables need to be predicted. Additionally, the algorithm leverages the simplified fixed Hessian method, which enhances computational efficiency and ensures that it can be employed in privacy-preserving scenarios through fully homomorphic encryption. The performance of the multi-output LFFR has been evaluated on various datasets, demonstrating its effectiveness and robustness in handling non-linear regression tasks while maintaining data privacy.

\bibliography{HE.LFFR}
\bibliographystyle{apalike}

\end{document}